\documentstyle[12pt,epsfig]{article}

\newcommand{\be}{\begin{equation}}
\newcommand{\ee}{\end{equation}}
\newcommand{\bd}{\begin{displaymath}}
\newcommand{\ed}{\end{displaymath}}
\newcommand{\ba}{\begin{eqnarray}}
\newcommand{\ea}{\end{eqnarray}}
\newcommand{\bi}{\begin{itemize}}
\newcommand{\ei}{\end{itemize}}

\newcommand{\sbbox}[1]{\mbox{\scriptsize\bf $#1$}}

\newlength{\baselineskipsave}
\newlength{\blss}
\def\dsl{\raise.15ex\hbox{/} \kern-.57em\partial}
\def\psl{\raise.15ex\hbox{/} \kern-.57em p}
\def\ksl{\raise.15ex\hbox{/} \kern-.57em k}
\def\qsl{\raise.15ex\hbox{/} \kern-.57em q}

\begin{document}

\title{
Inverse Symmetry Breaking on the lattice: an accurate MC study }

\author{G. Bimonte$^{b}$, D. I\~niguez $^{a}$, A. Taranc\'on  $^{a}$ and
C.L.Ullod $^{a}$ }
\bigskip
\maketitle

\begin{center}
{\it a)  Departamento de F\'{\i}sica Te\'orica, Facultad de Ciencias,\\
Universidad de Zaragoza, 50009 Zaragoza, Spain \\
\small e-mail: \tt david, tarancon, clu@sol.unizar.es} \\
{\it b)  Dipartimento di Scienze Fisiche, Universit$\grave{a}$ di Napoli,\\
Mostra d'Oltremare, Pad.19, I-80125, Napoli, Italy \\
\small e-mail: bimonte@napoli.infn.it\tt } \\
\end{center}
\bigskip

\begin{abstract}
{

We present here a new MC study of ISB at finite temperature in a $Z_2\times
Z_2$ $\lambda\phi^4$ model in four dimensions. The results of our
simulations, even if not conclusive, are favourable to ISB. Detection of
the effect required measuring some critical couplings with six-digits
precision, a level of accuracy that could be achieved only by a careful use
of FSS techniques. The gap equations for the Debye masses, resulting from
the resummation of the ring diagrams, seem to provide a qualitatively
correct description of the data, while the simple one-loop formulae appear
to be inadequate.}

\end{abstract}



\medskip
DFTUZ preprint 98/27~~~~~Napoli preprint 7/99~~~~~hep-lat/9903027

\newpage

\section{Introduction}

In 1974, in his classic paper on thermal field theory, Steven Weinberg,
quoting S. Coleman, reported on the unusual behavior of certain particle
theory models at high temperatures. Using the example of a simple
$O(N_1)\times O(N_2)$ scalar $\phi^4$ model, he showed that it is possible
to induce spontaneous symmetry breaking at arbitrarily high temperatures.
He named this funny phenomenon Inverse Symmetry Breaking (ISB) or
Symmetry-Non-Restoration (SNR), depending on whether the ground state was
disordered or ordered at zero temperature. In order to make this seemingly
paradoxical statement easier to accept, he observed that in Nature there is
a substance, the Rochelle salt, that does exhibit this sort of inverse
behavior. Its crystals are orthorhombic below the lower Curie point
(-$18^{~0}C$), and monoclinic above it, which means that the crystal has a
$\it larger$ symmetry group at the $\it lower$ temperature! Of course,
Weinberg's statement was much stronger than this example, for he claimed
that order could persist for ever, no matter how high the temperature,
while the Rochelle salt, if heated enough, eventually melts.

In any case, ISB and SNR would have probably remained relegated among the
amenities of field theory, if it were not for the fact that this mechanism
can be implemented in realistic particle models and then applied to
important cosmological issues related to the thermal history of the Early
Universe. This is not the right place to discuss these very interesting
applications of ISB and SNR, but to give the reader an idea of the number
and relevance of the issues that have been treated , we just make a list
and quote some references. Following the historical order, ISB and SNR were
applied to: the breaking of CP symmetry \cite{ms}, the monopole problem
\cite{sal}, baryogenesis \cite{dod}, inflation \cite{lee}, the breaking of
P, strong CP and Peccei-Quinn symmetries \cite{dva} and finally
supersymmetry \cite{rio}.

The potential importance of ISB and SNR for cosmology induced several
authors to study these phenomena more accurately, than was done in
Weinberg's original paper. His analysis relied, in fact, on a simple
one-loop computation, and so it was natural to question if higher order
corrections would have changed the scenario. This was not an idle question
for at least two reasons. On one side, it is well known that at high
temperature the reliability of perturbation theory is questionable, because
powers of the temperature can compensate for powers of the coupling
constants, spoiling the expansion. On the other side, and maybe more
importantly, when one tries to induce ISB or SNR in realistic particle
models, one is generally forced to consider rather large couplings in the
scalar sector, in order to overcome the action of the gauge fields, which
always work in favour of symmetry restoration at high temperature, and thus
it becomes important to explore the robustness of ISB and SNR with respect
to the coupling strengths.

The problem was attacked using a variety of approaches: large $N$
expansions \cite{fuj}, the Gaussian effective potential \cite{haj}, the
constraint effective potential \cite{fuj2}, gap equations \cite{bim}, the
average effective action \cite{roos} and recently chiral effective
Lagrangians \cite{gave}. The results were contradictory: some authors
concluded that ISB and SNR are artifacts of the one-loop approximation,
while others confirmed the existence of these phenomena, but observed that
higher order-corrections have the effect of reducing the size of the
parameter region for which they can occur.

This state of things motivated us to start a program of MC simulations to
study ISB, which seemed to us the only way to  carry a fully
non-perturbative analysis of this problem. We chose the simplest model that
can exhibit ISB, namely a $Z_2 \times Z_2$ two-scalars $\phi^4_d$ theory.
In the first run, we simulated the model in three dimensions \cite{bitu}
and afterwards we followed with four \cite{bim1}. In both cases we found no
sign of ISB, despite the fact that, for the values of the coupling
constants that were simulated, the one-loop conditions for ISB were
strongly satisfied. Recently, the authors of ref.\cite{jan} examined a $Z_2
\times O(4)$ two-scalar $\phi^4$ model, in four dimensions, and claimed
to have found a clear evidence that ISB was taking place.

In this paper we present the results of a new series of simulations of our
$Z_2 \times Z_2$ model. The strategy that we have followed is essentially
the same of our previous paper \cite{bim1}. Our two scalar model depends on
five parameters, three coupling constants $\lambda_1,\lambda_2$ and
$\lambda$, and two hopping parameters $\kappa_1$ and $\kappa_2$. We fixed
once and for all the values of the coupling constants: with respect to
\cite{bim1}, this time we enforced the perturbative conditions for ISB (in
the direction of the field $\phi_2$) much more strongly and at the same
time we took smaller values in order to be closer to the perturbative
region. We then studied the phase diagram as a function of the two hopping
parameters. In the $(\kappa_1,\kappa_2)$ plane we found transition lines of
first and second order, but since we were interested in ISB, we focused our
efforts on the critical line for the breaking of the field $\phi_2$. This
line is roughly parallel to the $\kappa_1$ axis and the aim was to
determine in what direction it shifts, when the temperature is increased:
ISB requires that, for $T>0$, it shifts towards smaller values of
$\kappa_2$, at least in a neighbourhood of the scaling region.

As it is well known, finite temperatures are simulated by lattices with a
finite extension $N_t$ in one direction, the temperature being $T=1/(N_t
a)$, with $a$ the lattice spacing. What made the simulations very hard is
the extreme smallness of the effect that we show in this paper: even
at the highest temperature, which corresponds to $N_t=2$, in order
to detect the shift
reliably, it was necessary to measure the critical values of $\kappa_2$,
both for $T=0$ and for $T>0$, with six significant digits! As it is well
known, the only safe way to achieve such a huge precision is via an
accurate analysis of Finite Size Scaling (FSS). This required large
lattices and tremendous statistics: in one case, for example, we simulated
a $20^4$ lattice, with $8\times 10^6$ iterations.

For $N_t=2$, the direction of the shift was favourable to ISB, but we
cannot consider this result as conclusive. Due to the difficulty of the
measurements, we could not simulate larger values of $N_t$, as it is
necessary in order to make sure, via scaling analysis, that things would go
in the same way for all $N_t$'s.

As a check, we compared the theoretical predictions with the MC value of
the critical temperature for which we observed ISB. While the simple
one-loop estimate is grossly incorrect, we found a reasonable agreement
with the value obtained from the gap equations, which result from resumming
the ring diagrams of the perturbative series \cite{dol}. These equations
seem to give a qualitatively correct description of the MC data, and
explain as well why we did not find ISB in our previous simulations in
\cite{bim1}.

We close this introductory Section with a plan of the paper: in Section 2
we introduce our lattice model and discuss its phase diagram at $T=0$. In
Section 3 we review the perturbative picture of ISB, while in Section 4 we
discuss in detail our strategy to study ISB on the lattice. In Section 5 we
discuss how the results of the simulations  compare with the theoretical
predictions, while Section 6  contains a discussion of the recent MC
studies in \cite{bim1} and \cite{jan}. Finally, Section 7 contains our
concluding remarks.

\section{The model and its lattice formulation}

We consider the theory for two real scalar fields in $4$ euclidean
dimensions, described by the bare (euclidean) action:
\be
S=\int d^4 x
\Big\{ \sum_{i=1,2}\Big[\frac{1}{2}
(\partial_{\mu}\Phi_i^{(0)2}) + \frac{1}{2}\overline{m}_i^{(0)2}
\Phi_i^{(0)2} + \frac{g_i^{(0)}}{4!} \Phi_i^{(0)4}\Big]+
\frac{1}{4}
g^{(0)}\Phi_1^{(0)2} \Phi_2^{(0)2} \Big\}.
\label{bact1}
\ee
What will be essential for ISB and SNR, in the above action the quartic
bare coupling $g^{(0)}$ can be {\it negative}. If $g^{(0)}>0$ the condition of
boundedness from below of the potential is satisfied if:
\ba
&&g_1^{(0)}> 0,~~~g_2^{(0)}> 0,\nonumber \\
&&g_1^{(0)}\; g_2^{(0)} > 9\; g^{(0)2}.
\label{cotas}
\ea

When regularized on an infinite four-dimensional cubic
lattice of points
$\Omega$ with lattice spacing $a$, the above action is replaced
by its discretized version
\ba
S_L&=& \sum_{x \in \Omega}   a^4 \Big\{ \sum_{i=1,2}\Big[
\sum_{\mu}\frac{1}{2}
(\Delta^{(a)}_{\mu}\Phi_{i,L})^2(x)+\frac{1}{2}
\overline{m}_i^{(0)2}\Phi_{i,L}^2(x)+
\frac{g_i^{(0)}}{4!}\Phi^4_{i,L}(x) \Big]+ \nonumber\\
&& \;\;\; \;\;\;\;\;\;\;\;\; + \frac{g^{(0)}}{4}\; \Phi_{1,L}^2(x)
\Phi_{2,L}^2(x)\Big\},\label{lat1}
\ea
where $\Delta^{(a)}_{\mu} $ is the lattice derivative operator in the
direction $\mu$:
\be
\Delta^{(a)}_{\mu}\Phi_{i,L}(x)\equiv \frac{\Phi_{i,L}(x +
\hat{\mu} a)-\Phi_{i,L}(x)}{a}\;.\nonumber
\ee
We find it convenient to measure all dimensionful quantities in (\ref{lat1})
in units of the lattice spacing; thus we define:
\be
\phi_{i,L}(x) \equiv a \Phi_{i,L}(x),~~~
m_i^{(0)2} \equiv a^2 \overline {m}_i^{(0)2}~~.\label{BBpar}
\ee
In terms of the dimensionless quantities the lattice action now reads:
\ba
S_L&=& \sum_{x \in \Omega}  \Big\{ \sum_{i=1,2}\Big[
\sum_{\mu}\frac{1}{2}
(\Delta^{(1)}_{\mu}\phi_{i,L})^2(x)+\frac{1}{2} m_i^{(0)2}\phi_{i,L}^2(x)+
\frac{g_i^{(0)}}{4!}\phi^4_{i,L}(x) \Big]+ \nonumber\\
&&\;\;\;\;  \;\;\;\;\;\;+ \frac{g^{(0)}}{4}\; \phi_{1,L}^2(x)
\phi_{2,L}^2(x)\Big\}\;,\label{lat2}
\ea
The standard lattice notation is obtained with a further redefinition
of the fields and couplings in (\ref{lat2}) according to:
\be
\phi_{i,L}(x) = {\sqrt \kappa_i}
\phi_{i,\sbbox{r}},~~~g_i^{(0)}=\frac{24 \lambda_i}{\kappa_i^2},~~~
m_i^{(0)2}=2\frac{1-2\lambda_i -4 \kappa_i}{\kappa_i}~~
\ee
and
\be
g^{(0)}=\frac{4 \lambda}{\kappa_1\kappa_2}.
\ee
After these redefinitions we get our final form of the lattice action
\ba
S_{L}&&=  \sum_{\sbbox{r} \in {\bf Z^4}}\Big\{ \sum_{i=1,2} \Big[
       -\kappa_i  \sum_{\mu}\phi_{i,\sbbox{r}} \phi_{i,\sbbox{r}+
\hat{\sbbox{\mu}}}+
       \lambda_i(\phi^2_{i,\sbbox{r}}-1)^2+\phi^2_{i,\sbbox{r}} \Big]+
\nonumber\\
&&~~~~~~~~~~~~~~~~~~~+\lambda\phi^2_{1,\sbbox{r}}\phi^2_{2,\sbbox{r}}\Big\}
\label{bbfin}.
\ea

For generic values of the parameters, the action has a ${\bf Z}_2 \times
{\bf Z}_2$ symmetry which can be spontaneously broken. The model is
expected to have only one fixed point, the Gaussian one, corresponding to
$m_1^{(0)2}=m_2^{(0)2}=g_1^{(0)}= g_2^{(0)}=g^{(0)}=0$ (or
$\kappa_1=\kappa_2=1/4$ and $\lambda_1=\lambda_2=\lambda=0$), which has a
null attractive domain in the infrared. In $4-\epsilon$ dimensions there
exist five more fixed points:

\noindent
1) the Heisenberg fixed point, for
$a^{\epsilon}g_1^{(0)}=a^{\epsilon}g_2^{(0)}= 3 a^{\epsilon}g^{(0)}=g^*_H$
and $m_1^{(0)2}=m_2^{(0)2}=m^{2*}_H < 0$, where the symmetry of the model
is enhanced to $O(2)$.

\noindent
2) the three Ising fixed points, for $g^{(0)}=0$ and
$a^{\epsilon}g_i^{(0)}=g_I^{*}$, $m_i^{(0)2}=m^{2*}_I$ for some $i$, where
the model splits into two independent $\phi^4_4$ models.

\noindent
3) the Cubic fixed point, for
$a^{\epsilon}g_i^{(0)}=a^{\epsilon}g^{(0)}=g_I^{*}/2$,
$m_i^{(0)2}=m^{2*}_I$, which again splits into two independent  $\phi^4_4$
models, after a $\pi/4$-rotation of the fields.

\begin{figure}[t!]
\epsfig{figure= 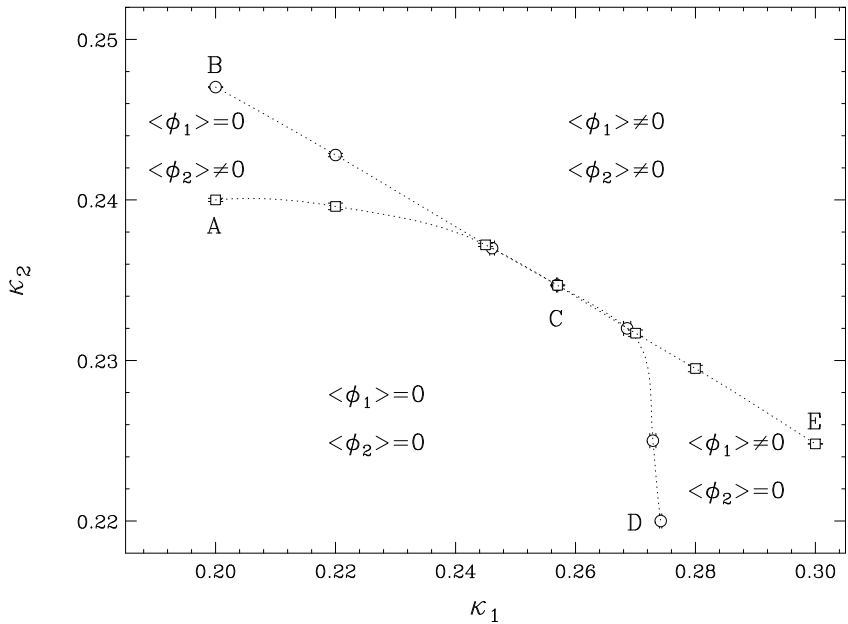,angle=90,width=320pt}
\caption{Phase diagram at $\lambda_1=0.3375,~\lambda_2=0.01125,
~\lambda=-0.112$.}
\label{phdi}
\end{figure}

The phase diagram of the model (\ref{bbfin}) (at $T=0$), for fixed values
of $\lambda_1$, $\lambda_2$ and $\lambda$, is shown in Fig. 1.

There are four distinct phases separated by four critical lines A, B, D and
E: the disordered one corresponds to the lower left corner of the picture,
while the upper right corner represents the totally ordered phase. The
wedges in between them are partially ordered phases: in the left wedge the
field $\phi_2$ is ordered, while $\phi_1$ is disordered; in the right wedge
it occurs the contrary.

If there was no interaction between $\phi_1$ and $\phi_2$,  we would have
two independent $\phi^4$ models and then the critical lines would all be of
second order. The lines A and E would form a straight line parallel to the
$\kappa_1$ axis, and similarly the lines B and D would form a straight line
parallel to the $\kappa_2$ axis. The presence of the interaction bends them
as can be seen in the Figure. We managed to determine the order of the
critical lines only at some distance from the region C, where the four
lines merge together. The critical line A is the only one that we studied
accurately: away from C, it is of second order and we observed for it
Gaussian critical exponents, as it had to be expected from triviality.
As for the
remaining critical lines, there is a clear evidence that D is of second
order, while B and E appear to be of first order. Close to C, things become
unclear: while on small lattices it becomes impossible to distinguish the
various phase transitions, on large ones the onset of strong
metastabilities prevented us from getting any indications at all.

Understanding the phase diagram in the neighbourhood of C is of course
extremely important. An important issue to address is whether the four
critical lines actually meet at a single point, a double critical point
that we will call ${\cal C}$, and what is its order.

Indeed, the indications coming from perturbation theory suggest that, when
$g^R$ is {\it negative}, as required for ISB or SNR to occur, ${\cal C}$,
if it exists, should be a first order critical point. This can be seen by
integrating the one-loop renormalization group equations (RGE), in the
scaling limit of large correlation lengths $\xi_i$ for both fields
$$
\xi_i=c_i \;\xi\;,\;\;\;\;\xi  \gg 1,
$$
where $c_i$ are constants. In this limit the RGE of our model read:
\ba
\frac{dg_1^R}{dt}&=&-\frac{3}{16 \pi^2}[( g_1^R)^2+ (g^R)^2]\;,\nonumber\\\
\frac{dg_2^R}{dt}&=&-\frac{3}{16 \pi^2}[( g_2^R)^2+ (g^R)^2]\;,\nonumber\\
\frac{dg^R}{dt}&=&-\frac{1}{16 \pi^2}(g_1^{R}+ g_2^R+4 g^{R})g^R\;,\label{rge}
\ea
where $t=\log \xi$. Numerical integration of these equations shows that, if
one starts from a set of initial values for the renormalized couplings with
$g^R$ negative, $g_1^R$ and $g_2^R$ are driven by $g^R$ towards {\it
negative} values in a finite interval of $t$, something that is generally
believed to indicate a {\it first-order} phase transition. The fact that
for negative values for the coupling $g^R$ the double critical point really
is of first order, is shown in ref.\cite{born}, where the phase diagram of
our model is studied, using a suitable truncation of the exact
non-perturbative flow equations, for the particular case
$g_1^{(0)}=g_2^{(0)},\;m_1^{(0)2}=m_2^{(0)2}$, when an extra $\phi_1
\leftrightarrow \phi_2$ symmetry is present (this case is not directly related to
ISB or SNR because this extra symmetry excludes a priori the possibility of
these phenomena).

If this picture of the double phase transition for $g^R<0$ turns out to be
correct, one may question whether in this case such a model can be used at
all as an elementary particles theory. The answer to this question depends
on the actual strength of the first-order phase transition. The fact that
the correlation lengths of one or both fields remain finite near $\cal C$
is not necessarily a problem per se. In fact, even if $\cal C$ were of
second order for both fields, triviality would prevent one from taking the
correlation lengths to infinity anyway. Thus, the relevant question is to
see if the correlation lengths become sufficiently large near $\cal C$, for
the presence of the lattice cut-off to become negligible in low-energy
amplitudes. This in turn depends on the absolute magnitude of the coupling
constants. For small couplings, $\cal C$ should be only weakly first order
and one should be able to achieve large correlation lengths for both
fields. In realistic particle models for ISB and SNR the problem may be
more serious: the presence of gauge couplings, that work in favour of
symmetry restoration, usually forces one to take large negative couplings
$g^R$, for ISB or SNR to occur \cite{bim}, and this can then give rise to
rather strong first order phase transitions in the scalar sector, thus
spoiling the model.

\section{High-T perturbation theory}

In this Section we briefly review the predictions of perturbation theory on
the high-temperature behavior of the model (\ref{bact1}). We shall see
that, for certain choices of the couplings, the symmetric vacuum of the
theory becomes unstable at sufficiently high temperatures. In order to
explore this phenomenon, we need to compute the leading high-$T$
contribution to the second derivatives of the effective potential in the
origin, or equivalently the 1PI 2-point functions for zero external momenta
$p$. A standard one-loop computation (in the imaginary time formalism)
\cite{kap} in our model gives the result:
\ba
M_1^2(T) &\equiv& -\Gamma^{(2)}_1(p=0,T)= m_1^2 +
\frac{T^2}{24}\left[g_1^R h\left(\frac{m_1}{T}\right)+ g^R
h\left(\frac{m_2}{T}\right)\right]~~,\nonumber\\ M_2^2(T) &\equiv&
-\Gamma^{(2)}_2(p=0,T)= m_2^2 +  \frac{T^2}{24}\left[g_2^R h\left(\frac{m_2}{T}\right)+ g^R
 h\left(\frac{m_1}{T}\right)\right]~~,\label{selft}
\ea
where $m_i^2$ represent the renormalized masses at $T=0$ and $h(x)$ is the
function:
\be
h(x)=\frac{6}{ \pi^2}\int_0^{\infty} dy\;
\frac{y^2}{(x^2+y^2)^{1/2}(e^{(x^2+y^2)^{1/2}}-1)}~.
\ee
It is clear that $h(x)$ is positive, monotonically decreasing and that it
approaches zero when $x\rightarrow
\infty$. For small values of $x$, $h(x)$ has the asymptotic
expansion:
\be
h(x)=1-\frac{3}{ \pi}x-\frac{3}{2
\pi^2}x^2\left(\ln\frac{x}{4\pi}+\gamma-\frac{1}{2}\right),
\ee
$\gamma$ being the Euler constant. In the limit of very high-temperatures
$m_i/T \ll 1$, eqs.(\ref{selft}) reduce to:
$$
M_1^2(T) =m_1^2 + \frac{T^2}{24} (g_1^R + g^R)~~,
$$
\be
M_2^2(T) =m_2^2 + \frac{T^2}{24}(g_2^R + g^R)~~.\label{onel}
\ee
Assuming for simplicity that the bare couplings are so small that they can
be identified with the renormalized ones, and recalling that the coupling
$g^{(0)}\approx g^R$ can be {\it negative}, it is easy to see that for
$g_1^R/g_2^R$ sufficiently large, the range (\ref{cotas}) of stability for
the potential includes values of $g^R$ such that, say, $g_2^R + g^R$ is
{\it negative} (it can be proven easily form eqs.(\ref{cotas}) that one
cannot have $g_1^R+g^R<0$ at the same time). In this case we see from
eq.(\ref{onel}) that $M_2^2(T)$ becomes negative at sufficiently high
temperatures, irrespective of its value at $T=0$. This is the essence of
the phenomena of SNR and ISB: in multiscalar models some Debye masses can
become negative and so one can have spontaneous symmetry breaking at
arbitrarily high temperatures. We now focuse on ISB, the phenomenon to be
explored in this paper: in this case we take $m_1^2>0, \;m_2^2>0$, so that
the vacuum is disordered at $T=0$, and again assume that $g_2^R + g^R<0$.
If one starts increasing the temperature, it will happen that, at a certain
critical temperature $T_c$, the field $\phi_2$ will undergo spontaneous
symmetry breaking and will stay broken at all higher temperatures. The
field $\phi_1$, instead, will remain disordered. The critical temperature
$T_{c}$ can be estimated from eq.(\ref{onel}) to be:
\be
\frac{T_{c}}{m_2}=\sqrt{-\frac{24}{g^R+g_2^R}}\;\;.\label{naive}
\ee
Even though they capture the main features of ISB and SNR, at a closer
inspection, eqs.(\ref{onel}) and (\ref{naive}) appear unsatisfactory in two
respects. First of all, the condition for the high-T instability in, say,
the $\phi_2$ direction, $g_2^R+g^R<0$,  as deduced from the second of
eqs.(\ref{onel}), does not involve $g_1^R$. Since the bound (\ref{cotas}),
together with the condition $g_2^R+g^R<0$, typically implies a hierarchy
among the couplings $g_1^R
\gg |g^R| \ge g_2^R$, it is clear that next-to-leading order corrections
involving $g_1^R$ can produce significant corrections to the lowest order
result. The second fault of eqs.(\ref{onel}) and (\ref{naive}) is the most
important for our study of ISB. It is the fact that the estimate
(\ref{naive}) of the critical temperature does not involve $m_1^2$: this
looks physically incorrect, for one can imagine that for $m_1\gg m_2$ the
field $\phi_1$ should decouple from $\phi_2$ for temperatures $T \ll m_1$
and so ISB or SNR should not occur before the scale $m_1$ is reached. Both
limitations of the one-loop picture can be overcome by performing the
resummation of the so called ring diagrams.

As it is well known \cite{dol}, these diagrams represent the dominant
next-to-leading corrections at high temperatures and their inclusion leads
to a more accurate description of both the phase transition and the high
temperature behavior of the thermal masses. This infinite resummation leads
to self-consistent gap equations for the Debye masses $M_i^2(T)$, which in
our case read
\ba
x_1^2 &=&\frac{m_1^2}{T^2}+  \frac{g_1^R}{24}  h(x_1)+ \frac{g^R}{24}
h(x_2)~~,\nonumber\\ x_2^2&=&\frac{m_2^2}{T^2}+ \frac{g_2^R}{24} h(x_2)+
\frac{g^R}{24} h(x_1)~~,\label{gapeq}
\ea
where we defined $x_i=M_i(T)/T$. For any given temperature $T$, the
symmetric vacuum is stable if the above equations admit real (positive)
solutions for $x_i$. At sufficiently low temperatures, this is clearly the
case if we take $m_1^2>0,\;m_2^2>0$. One can now gradually increase $T$ and
follow the evolution of $x_1$ and $x_2$: if one of them, say $x_2$,
approaches zero at $T=T_c$, and if for $T>T_c$ the gap equations do not
admit anymore real solutions, one can argue that for $T>T_c$ there will be
ISB in the $\phi_2$ direction. It is clear that this can happen only if
$g^R$ is negative and its absolute value is large enough in comparison with
$g_2^R$. A detailed study of the conditions required is given in the first
of refs.\cite{bim}, where it is shown that the parameter region for which
ISB or SNR occur is indeed smaller than that predicted by the naive
eqs.(\ref{onel}). An analogous result was later confirmed by studies based
on different non-perturbative techniques \cite{roos}.

Assuming now that at high temperatures there develops an instability in the
$\phi_2$ direction, a better estimate of the critical temperature $T_c$,
than that given by the one-loop equations (\ref{naive}), can be obtained by
setting $x_2=0$ in eqs.(\ref{gapeq}) and then solving the system:
\ba
x_1^2 &=& \frac{m_1^2}{T_{c}^2} +  \frac{1}{24}(g_1^R  h(x_1)+
{g^R})~~,\nonumber\\ 0 &=&\frac{m_2^2}{T_{c}^2}+ \frac{1}{24}(g_2^R + g^R
h(x_1))~~,\label{crit}
\ea
where we have used $h(0)=1$. Upon comparing the second of eqs.(\ref{crit})
with the second of eqs.(\ref{onel}), it is clear that the inclusion of the
ring diagrams pushes the phase transition to higher temperatures. This is
so because the negative coupling $g^R$, which is the cause of ISB, comes
with the factor $h(x_1)$, which is always less than one. Since $x_1$ is
larger then $m_1^2/T_c^2$, as it is clear from the first of
eqs.(\ref{crit}), we see that a large mass $m_1^2$ will make ISB harder to
achieve. In fact, the influence of $m_1^2$ can be so strong as to cause the
disappearance of the symmetry-breaking phase transition predicted by the
one-loop formula. As we shall explain in greater detail in Sec. 6, this
effect probably explains why we did not observe ISB in our previous work
\cite{bim1}.

\section{ISB on the lattice}

We saw in the previous Section that, for $g_2^R+ g^R$ sufficiently
negative, continuum perturbation theory predicts ISB for the $\phi_2$
field. Even if the mass parameters are chosen in such a way that the ground
state is disordered at $T=0$, the field $\phi_2$ should develop a
non-vanishing vev above some critical temperature $T_c$ and should remain
ordered at all higher temperatures. The field $\phi_1$, instead, should
remain disordered.

In principle it is very simple to study ISB on the lattice. As it is well
known, finite temperatures are simulated by lattices with a finite
extension in the ``time" direction (and infinite extension in the remaining
``space" directions, as required by the thermodynamic limit), the
temperature being related to the number $N_t$ of sites by the relation:
\be
T=\frac{1}{N_t a}\;,
\ee
with $a$ the lattice spacing. So, in order to study ISB, one should simply
check how the critical lines of the $T=0$ phase diagram, shown in Fig.1,
shift in the $(\kappa_1,
\kappa_2)$ plane (for fixed values of $\lambda_1$, $\lambda_2$ and
$\lambda$) as a function of the temperature, which means concretely as a
function of $N_t$ (in principle, one could vary the temperature also by
varying the lattice spacing in the time direction, but we have not used
this method). Since we are specifically interested in the behavior of the
field $\phi_2$, it is sufficient to focuse our attention on the critical
line of this field, the line A-E of Fig.1. If, for the selected values of
$\lambda_1$, $\lambda_2$ and $\lambda$, ISB occurs, it should happen that,
when the temperature is increased, namely when $N_t$ is diminished, the
$\phi_2$-critical line penetrates more and more deeply in the disordered
region of the $T=0$ phase diagram, at least in a neighbourhood of the
double critical point ${\cal C}$. The effect should be clearest at the
highest temperatures, i.e. for the smallest values of $N_t$. To help the
reader visualize the situation, we have shown in Fig.2 the expected
$N_t$-behavior of the $\phi_2$-critical line, in case of ISB. Some comments
are in order.
\begin{figure}[t!]
\epsfig{figure= 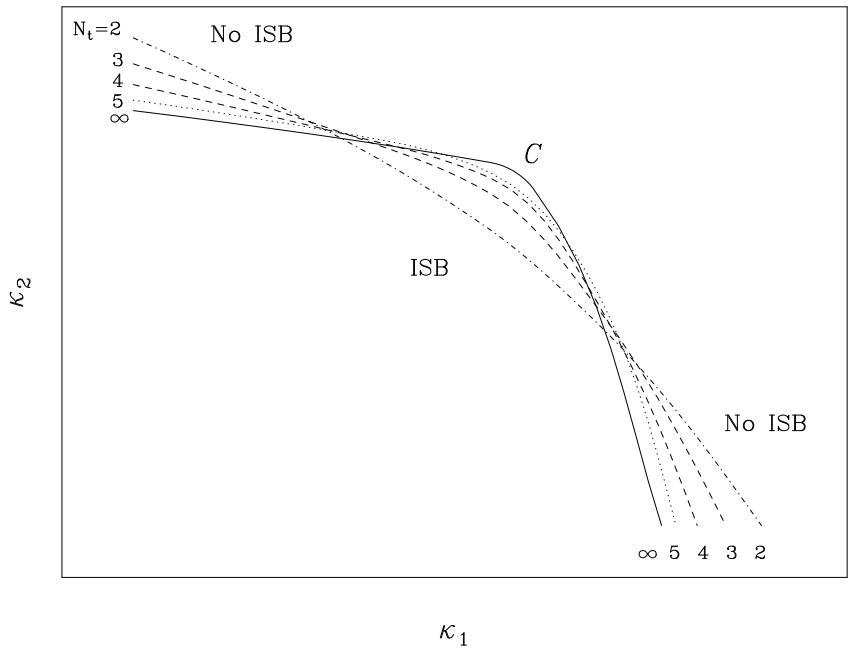,angle=90,width=320pt}
\caption{Expected (qualitative) temperature-behavior of the
critical line for the $\phi_2$ transition, in case of ISB. The region
studied by us is the one to the left of {\it C}.}
\label{evolucion}
\end{figure}

We expect that far from the double critical point ${\cal C}$ the $N_t$
behavior of the $\phi_2$-critical line should always be the normal one,
namely the critical line should move towards greater values of $\kappa_2$
when $N_t$ is diminished. This conjecture looks reasonable, if one
considers that far from $\cal C$, the field $\phi_1$ should be far from
critical and its correlation length should become very small. This implies
that the influence of the field $\phi_1$ over the field $\phi_2$, near the
critical line of the latter, should be negligible and so the behavior of
$\phi_2$ should be that of a normal $\phi^4$ theory, where an increase of
temperature shifts the transition upwards. This means that ISB should be
visible, if it happens at all, only close enough to ${\cal C}$, in the
scaling region where the correlation lengths of both fields become large.
This explains why, in Fig. 2, we have drawn the critical lines for
$N_t=2,3,4,5$ below that of the $T=0$ theory ($N_t=\infty$) only near
${\cal C}$, while away from ${\cal C}$ they all lie above it.

Such a behavior poses a very serious problem: imagine that for the few
values of $N_t$ simulated, one has found a result like that shown in Fig.
2, which supports ISB. How can one convince oneself that nothing will
change for greater values of $N_t$? One could suspect that for $N_t$ large
enough the critical line might pass entirely above that of the $T=0$
theory, thus leading to the disappearance of ISB. In practice, it is then
very important to have a criterion to decide if, whatever behavior is
observed with the few $N_t$'s that are simulated, it will not change when
going to higher $N_t$'s. The common practice to address this question is to
search for scaling in the $N_t$ dependence of some observable: this is what
we did in \cite{bim1}, where, for a fixed value of $\kappa_1$ and several
values of $N_t$, we measured the critical values of $\kappa_2$,
$\kappa_{2}^c(N_t)$. The values we got were all greater than the critical
value $\kappa_2^c$ for $T=0$, a result against ISB (very likely, the value
of $\kappa_1$ used for the simulations in \cite{bim1} was too far from
${\cal C}$, for ISB to be visible. We were probably in the region of Fig. 2
on the left of ${\cal C}$, where the behavior of the $\phi_2$ critical line
becomes the ``normal" one. We shall have more to say on this in Sec. 6.) .
The good scaling of $\kappa_{2}^c(N_t)$ with $N_t$ convinced us that no
change would have occurred for larger $N_t$. We think that applying this
method to make sure that ISB survives for all values of $N_t$ would be very
difficult, because we believe that in order to observe ISB for a number of
values of $N_t$ large enough for a scaling analysis to be possible, one
needs look very close to the critical point ${\cal C}$, where the large
correlation lengths make it necessary to use very large lattices for a
reliable measurement of the critical couplings.

\begin{table}[t!]
\begin{center}
\begin{tabular}{ccccc}
\hline
$N_t$ & $N_s$ &\# runs& Thermalization sweeps & \# sweeps \\ \hline \hline
  2   &   6   &   1   &  $5\times 10^4$      & $4\times 10^6$   \\
  2   &   8   &   1   &  $10^5$              & $8\times 10^6$   \\
  2   &   10  &   1   &  $2\times 10^5$      & $ 10^7$          \\
  2   &   12  &   1   &  $3\times 10^5$      & $ 10^7$          \\
  2   &   16  &   2   &  $2\times 10^5$      & $8\times 10^6$   \\
  2   &   20  &   4   &  $8\times 10^4$      & $6.4\times 10^6$ \\
\hline
  6   &   6    &  1   &  $5\times 10^4$      & $4\times 10^6$   \\
  8   &   8    &  1   &  $10^5$              & $8\times 10^6$   \\
 10   &  10    &  1   &  $2\times 10^5$      & $4\times 10^6$   \\
 12   &  12    &  1   &  $3\times 10^5$      & $8\times 10^6$   \\
 16   &  16    &  4   &  $2\times 10^5$      & $9.6\times 10^6$ \\
 20   &  20    &  6   &  $2.4\times 10^5$    & $9.6\times 10^6$ \\
\hline
\end{tabular}
\end{center}
\caption{Statistics for the simulations on lattices of sizes
$N_t\times N_s^3$.}
\label{tablastat}
\end{table}

As a matter of fact, in the new series of simulations presented here, we
have been able to simulate only one value of $N_t$, $N_t=2$, and we needed
a very large statistics in order to detect reliably the shift of the
$\phi_2$ critical line.

Let us briefly explain the method we have followed, postponing the details
to the next two subsections.

The first part of the method is exactly the same as that of
ref.\cite{bim1}, but it is useful to review it here again. We fixed once
and for all a set of values for $\lambda_1,
\lambda_2$ and $\lambda$. This time, we took $\lambda_1=0.16875$, $\lambda_2=0.0001125$,
$\lambda=-0.00784$.  There are two important differences with
ref.\cite{bim1}:

\noindent
a) the new values of the couplings are smaller than the old ones
\cite{bim1}. This was done to be closer to the perturbative region, but it
increased tremendously the difficulty of the simulations, because the shift
of the critical line with $N_t$ was this time  much smaller than before;

\noindent
b) more importantly, the ratio $|\lambda|/\lambda_2$ is now much greater
than before (more or less 70 against 10), and so there is now a much
stronger push towards ISB.

After a quick analysis of the phase diagram at $T=0$, we selected a value
of $\kappa_1$, $\kappa_1=0.24$, to the left of $\cal C$, and then we
measured accurately the corresponding critical value $\kappa_2^c$ of
$\kappa_2$ along the $\phi_2$ critical line, always at $T=0$.

Near this point, we measured the $T=0$ values of the renormalized couplings
and the correlation lengths of both fields, for the same value of
$\kappa_1$, but for a value of $\kappa_2$ slightly less than $\kappa_2^c$,
in the disordered phase (the results of these measurements are given in
Table 2). The value of $\xi_1\simeq 1$ that we got is not very large, but
still acceptable for us to believe that we were probing the continuum
region. In any case, we could not achieve larger values for $\xi_1$,
because near to ${\cal C}$ we encountered strong metastabilities that made
any simulations impossible. We notice also that the value of $g_1^R$ is
quite large and not very accurate; on the contrary, what we think is the
essential, we see that $g_2^R$ and $|g^R|$ are both very small and that the
condition for ISB is implemented very strongly, since $|g^R|/g_2^R \simeq
10$. The one-loop formulae obviously predict that ISB should occur for
these values of the couplings, and it can be checked that the ratio
$|g^R|/g_2^R$ is large enough for ISB to survive also the inclusion of the
ring diagrams. This convinced us that we had found a good simulation
region.

Having performed all these measurements at $T=0$, we turned to finite
values of $N_t$. Due to the extreme smallness of the effect, we were able
to detect the shift of $\kappa_2^c$ (for $\kappa_1=0.24$) only for the
smallest value of $N_t$, $N_t=2$. Even in this case, the shift was of the
order of one part in hundred thousands and so larger values on $N_t$ look
out of reach. Since this shift turned out to be towards smaller values of
$\kappa_2$, our result can be considered as favourable to ISB.

In the next two Subsections, we describe the details of the numerical
simulations and how we managed to measure the critical values of $\kappa_2$
with the high accuracy that was needed.

\subsection{MC Simulation: determination of $\kappa_2^c$}

\begin{figure}[t!]
\epsfig{figure= 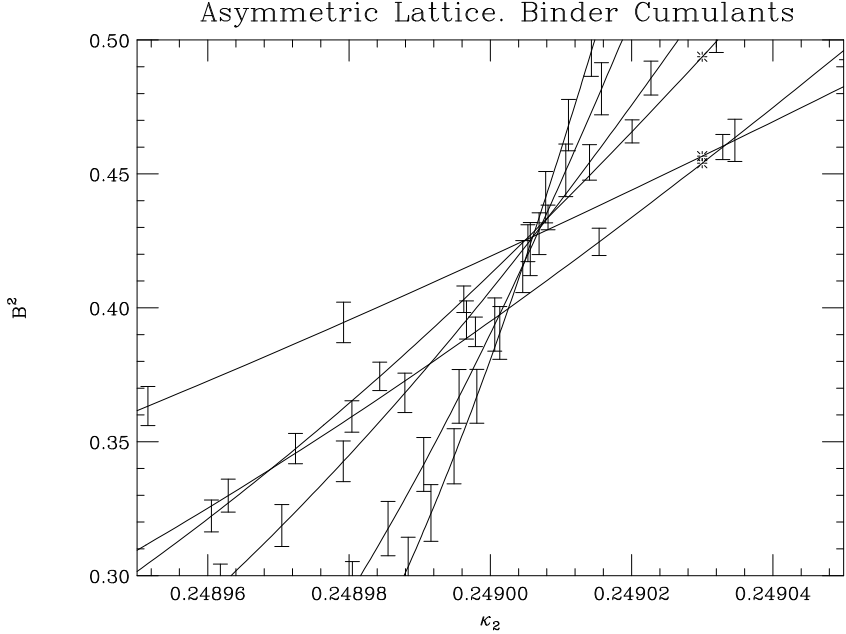,angle=90,width=320pt}
\caption{Binder cumulants for the $2\times N_s^3$ lattices ($N_s=6,8,10,12,16,20$)
at $\lambda_1=0.16875,~\lambda_2=0.0001125,~\lambda=-0.00784$ and
$\kappa_1=0.24$.}
\label{cortesnt2}
\end{figure}

For this new simulation, we used the Metropolis Algorithm (cluster was not
efficient), with $3$ hits and acceptance around $60\%$. We simulated
lattices of sizes $2\times N_s^3$ and $N_s^4$, with $N_s=6,8,10,12,16,20$.
In the new region of parameters the field fluctuations are strong and the
transitions for different lattice sizes are very close. We had to
accumulate very large statistics in order to obtain accurate results. In
Table \ref{tablastat} we summarize the statistics for every lattice: number
of runs, sweeps left for thermalization  and total number of sweeps for
every run.

As we said above, all our simulations were performed for a single value of
$\kappa_1$, $\bar{\kappa}_1$=$0.24$. A fast simulation on the $6^4$ lattice
gave us the approximate position, $\kappa_2=0.24903$, of the phase
transition of the field $\phi_2$ (the field $\phi_1$ remains disordered
across the transition). All the subsequent massive simulations were
performed at this point, and the results were extrapolated in a narrow
$\kappa_2$-interval around it by means of the Spectral Density Method
\cite{ferr} (SDM). The observable used to locate with high precision the
phase transition was the Binder cumulant:

\be
U_i(N_s,\kappa_1,\kappa_2)=\frac{3}{2}-
\frac{\langle M_i^4 \rangle}{2\langle M_i^2\rangle^2}~,\label{bin}
\ee
where $M_i$ stands for the magnetization of the field $\phi_i$.

\begin{figure}[t!]
\epsfig{figure= 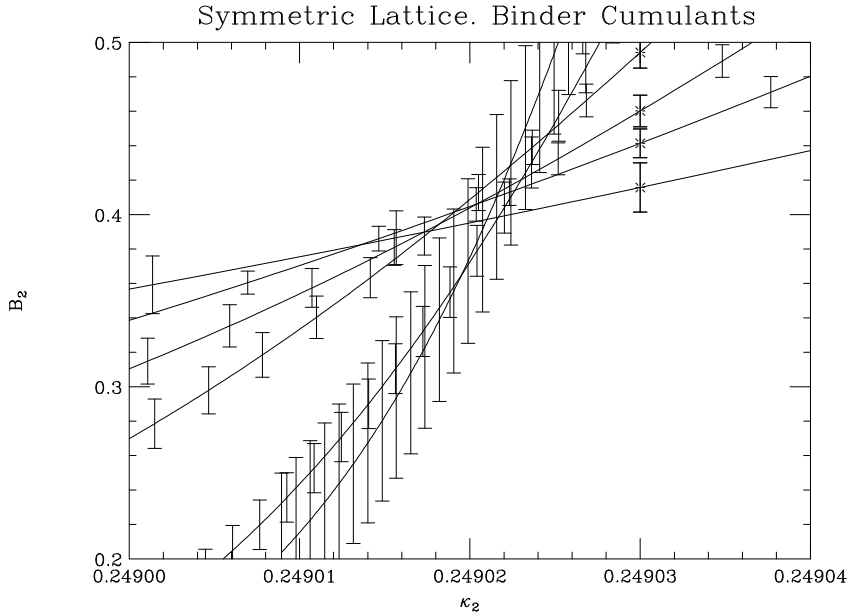,angle=90,width=320pt}
\caption{Binder cumulants for the $N_s^4$ lattices ($N_s=6,8,10,12,16,20$)
at $\lambda_1=0.16875,~\lambda_2=0.0001125,~\lambda=-0.00784$ and
$\kappa_1=0.24$.}
\label{cortessim}
\end{figure}

Figs.\ref{cortesnt2} and \ref{cortessim} show the values of $U_2$ obtained
by extrapolating the results of the simulations  by means of the SDM. Every
pair of curves ($N_s$ and $\hat{N_s}$) determines a crossing point, that
gives an estimate $\kappa_2^*(N_s,\hat{N_s})$ of the critical point
$\kappa_2^c$. These values do not suffer from finite size effects, except
for corrections to scaling, which can be parametrized and used to obtain
$\kappa_2^c$, according to the scaling law:

\be
\kappa^*_2(N_{s}, b N_{s})- \kappa_2^c =
\frac{1-b^{-\omega}}{b^{1/\nu}-1} N_s^{-\omega -1/\nu}~,\label{binsca}
\ee
where $\omega$ is the exponent for the corrections to scaling. In the
calculation of the error, due account was taken of the fact that the pair
crossings were not all independent of each other (out of the $N(N-1)/2$
crossings between $N$ curves, only $2N-3$ are independent of each other).

As for the exponents occurring in eq. (\ref{binsca}), we had to distinguish
the symmetric ($N_s^4$) lattices from the asymmetric ($2\times N_s^3$)
ones. In the case of the latter, since the scaling parameter is $N_s$,
while $N_t$ is fixed, according to the hypothesis of dimensional reduction
and universality, we used the exponents of the Ising model in {\it three}
dimensions, namely $\nu =0.63$ and $\omega=0.8$. As a check, we also
computed $\nu$ directly from the data obtaining values fully compatible
with the above one (see Fig. \ref{nu}).

In the case of the symmetric lattices, we used the mean field exponents
$\omega=0,\nu=1/2$. As seen from eq.(\ref{binsca}), these values imply
that, apart from logarithmic corrections, all the crossings should occur
for the same value of $\kappa_2$, and this is approximately what we found.
A better estimate of $\kappa_2^c$ was then obtained by setting $\omega \neq
0$ in eq.(\ref{binsca}) and taking the limit, for $\omega\to 0$, of the
corresponding values of $\kappa_2^c$. In the process, we observed a very
fast convergence.

\begin{figure}[t!]
\epsfig{figure= 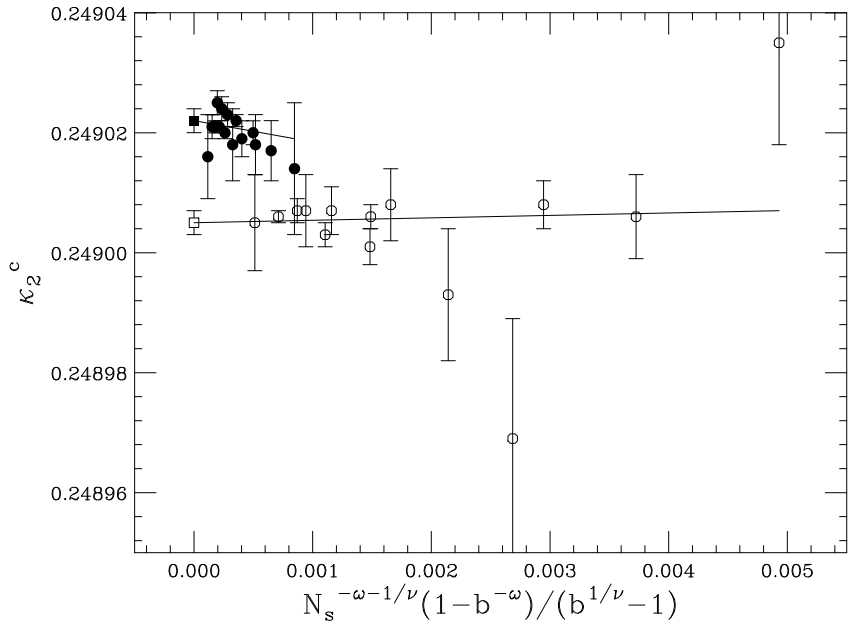,angle=90,width=320pt}
\caption{Fits to obtain $\kappa_2^c$ from the crossings
of the Binder cumulants. The results are $\kappa_2^c(T=0)=0.249022(2)$
(filled) and $\kappa_2^c(N_t=2)=0.249005(2)$ (empty).}
\label{fscall}
\end{figure}

The result of the extrapolation for $N_s \to \infty$ is shown in Fig.
\ref{fscall}. We got $\kappa_{2}^c(N_t=2)=0.249005(2)$ and
$\kappa_{2}^c(T=0)=0.249022(2)$. This means that the disordered region
diminishes when $T$ is increased, as ISB requires, at least for the case of
the highest temperature reachable on the lattice ($N_t=2$).

\subsection{MC Simulation: maximum of the Binder cumulant derivative}

\begin{figure}[t!]
\epsfig{figure= 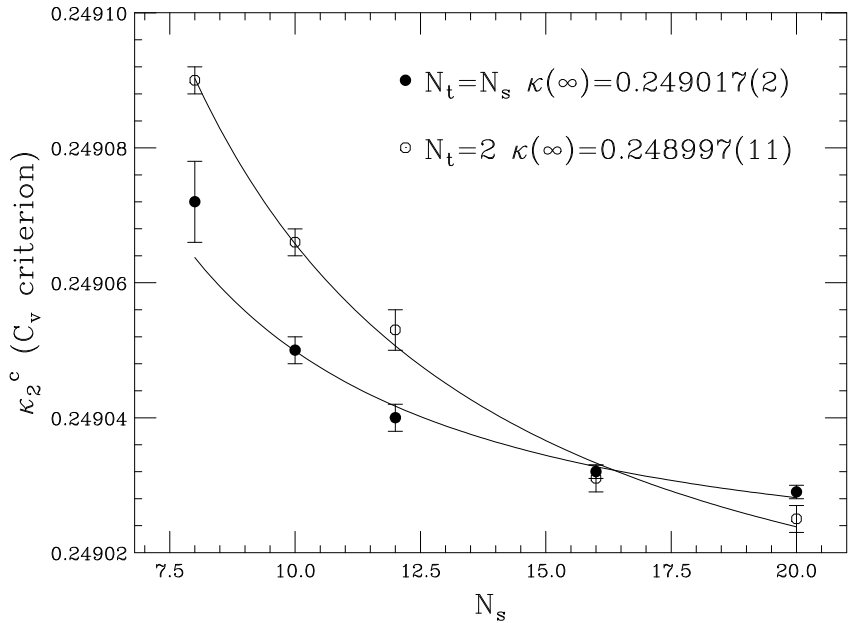,angle=90,width=320pt}
\caption{Fits to obtain $\kappa_2^c$
from the maximum of the Binder cumulant derivative. (The exponents used
were $\nu=0. 63,\omega=0.8$ for $N_t=2$ and $\nu=1/2$ for $N_t=N_s$).}
\label{maxcv}
\end{figure}

Due to the extreme smallness of the shift in the value of $\kappa_2^c$, we
found it opportune to compute the two values of $\kappa_2^c$ also in
another way, completely independent on the previous one. Indeed, for every
value of $N_s$, one can determine the point $\kappa_2^{**}(N_s)$ where the
derivative $\partial U_2/\partial
\kappa_2$ reaches its maximum value, namely $C_v$. It is well known that,
in the limit $N_s\rightarrow \infty$, $\kappa_2^{**}(N_s)$ approaches
$\kappa_2^c$ according to the scaling law (including corrections to
scaling):

\be
\kappa_2^{**}(N_s)-\kappa_2^c\propto N_s^{-1/\nu}(1-AN_s^{-\omega}) \,
\label{kappacv}
\ee
which allows us to compute $\kappa_2^c$. Fig.\ref{maxcv} shows the
corresponding fits. In this computation, we used the same critical
exponents discussed in the previous subsection. We obtained
$\kappa_2^c(N_t=2)=0.248997(11)$ and $\kappa_2^c(T=0)=0.249017(2)$, which
are fully compatible with the previous results.

\begin{figure}[t!]
\epsfig{figure= 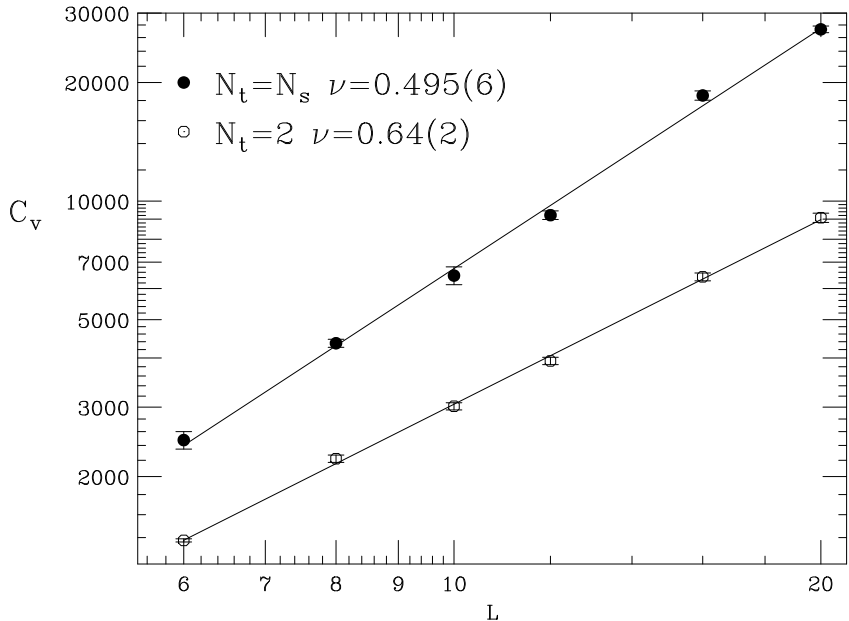,angle=90,width=320pt}
\caption{Fits to obtain $\nu$ by using the maximum of the
Binder cumulant derivative.}
\label{nu}
\end{figure}

As a final check, we computed $\nu$ from the maximum of $\partial
U_i/\partial \kappa_2$, $C_v$, which diverges with the lattice size as
$N_s^{1/\nu}$. Again, we obtained consistent results, as can be seen in
Fig. \ref{nu}.

In conclusion, this time, differently from what we reported in \cite{bim1},
we found that, when $N_t=2$, the phase transition of the field $\phi_2$
occurs for a value of $\kappa_2$ which is {\it smaller} than at $T=0$. The
two measures of $\kappa_2^c$ took a long time, because, for the small
values of $\lambda_2$ and $\lambda$ used now, the shift of the critical
$\phi_2$-line is extremely small and we had to measure both values of
$\kappa_2^c$ with six significant digits, a job that took us months of
computer time!

So, the results of our new simulations are favourable to ISB, but cannot be
considered as conclusive because they refer to a single value of $N_t$.

\section{Comparison with Perturbation theory}

As we said in Section 3, simulating only a finite number of $N_t$ values is
{\it per se} not sufficient to claim ISB,  for one cannot rule out the
possibility that the phenomenon would disappear for larger values of $N_t$.
In fact, the case here is even worse, since we have at our disposal just
one value of $N_t$, $N_t=2$. In order to increase confidence in our result,
we attempted a comparison of our MC data with the perturbative predictions.
While the one-loop formulae, eqs.(\ref{naive}), lead to an estimate of the
critical temperature far from the MC value, we found a reasonable agreement
with the predictions of the gap equations, eqs.(\ref{crit}).

\begin{table}[h!]
\begin{tabular}{|c|c|c|c|c|c|}
\hline
$N_s$ & $\xi_1$ & $\xi_2$ & $g_1^R$ & $g_2^R$ &  $g^R$\\
\hline
6 & 0.9581(2) & 10.78(1) & 26(6) & 0.0261(9) & -0.35(1) \\
\hline
8 & 0.9528(3) & 11.48(2) & 41(11) & 0.0296(15) & -0.29(1) \\
\hline
\end{tabular}
\caption{$T=0$ values of the correlation lengths and renormalized couplings for
 $\kappa_1=\bar{\kappa}_1\equiv 0.24$,
$\kappa_2=\bar{\bar{\kappa}}_2\equiv 0.2488$ (lattice sizes $N_s^4$).}
\label{table2}
\end{table}

We shall now explain how this comparison was done. Let $
p=(\bar{\kappa}_1,\bar{\kappa}_2)$ be the point of the $\phi_2$-critical
line for $N_t=2$, that we identified in our simulations. As we saw, $p$
belongs to the disordered phase of the $T=0$ phase diagram. If we knew the
$T=0$ values of the renormalized couplings and correlation lengths $\xi_1$
and $\xi_2$ (we measure them in units of the lattice spacing) at $p$, we
could get an estimate of the critical inverse temperature $N_t^c$ by
solving eq.(\ref{naive}) or eqs.(\ref{crit}), since
\be
\left(\frac{T_c}{m_i}\right)^{MC}=\frac{\xi_i}{N_t^c}~.\label{MC}
\ee
If perturbation theory works, we should get for $N_t^c$ a value close to
two.

This procedure, even if simple from the conceptual point of view, is
difficult to carry out: since $p$ is very close to the critical line of the
$T=0$ phase diagram, the correlation length $\xi_2$ is huge there (in our
case, we estimate it to be larger than 40). Thus a measure of the $T=0$
renormalized couplings and correlation lengths at $p$ requires 4d-lattices
with $N_s\geq 40$, which is just too much for computers available today.

In order to avoid the difficulties connected with exceedingly large
correlation lengths for the field $\phi_2$, we measured the renormalized
couplings and the correlation lengths (for $T=0$) at a point $\bar
{p}=(\bar{\kappa}_1,\bar{\bar{\kappa}}_2)$, having the same $\kappa_1$
coordinate as $p$, but with $\bar{\bar{\kappa}}_2$ slightly less than
${\bar{\kappa}}_2$, and such that $\xi_2$ was more or less ten (see Table
\ref{table2}). Because of time limitations, we did not carry out the
thermodynamic limit in this series of measures and contented ourselves with
simple estimates on rather small lattices. The quantitative considerations
that follow have to be taken, then, as simply indicative.

Afterwards, we computed the values of $g_1^R$, $g_2^R$, $g_R$ and of
$\xi_1$ and $\xi_2$ at $p$, by integrating a set of ($T=0$) RGE's in the
interval $\bar{\bar{\kappa}}_2<\kappa_2<\bar{\kappa}_2$. We evaluated the
$\beta$-functions to two-loops order, following a procedure strictly
analogous to that discussed in \cite{lush}. Of course in our case there was
a larger number of $\beta$ functions to be computed. Since the value of
$\xi_1$ at $\bar {p}$ was only roughly one, we included in our
$\beta$-functions the corrections to scaling upto one-loop order.

The reader might question the reliability of these perturbative RGE's,
since the value of $g_1^R$ at $\bar {p}$ is rather large and quite
uncertain. We believe that this is not a problem, for two reasons. The
first one is that the integration interval in $\kappa_2$ is very small. The
second, and most important, is less obvious: it turns out that $g_1^R$ does
not appear in any of the $\beta$ functions (at least to the two-loops order
that we have examined). Its absence can be understood by considering that,
since we are moving in the vertical direction in the $(\kappa_1,~\kappa_2)$
plane, $\xi_1$ remains finite (and practically constant).

After the numerical integration, it turned out that the running of $g_1^R$,
$g_2^R$, $g^R$ and $\xi_1$ in the tiny $\kappa_2$ interval considered was
completely negligible, while $\xi_2$ varied from 11 to something like 45.
Since the final value of $\xi_2$ was very sensitive to its initial value,
(and practically insensitive to the initial values of the other
parameters), we took advantage of our accurate knowledge of
$\kappa_2^c(T=0)$, and tuned the initial value of $\xi_2$ in such a way
that $\kappa_2$ would approach the MC value of $\kappa_2^c(T=0)$ in the
limit $\xi_2\rightarrow \infty$. The initial value of $\xi_2$ that we
obtained in this way, $\xi_2=11.85(5)$, is very close to the MC one, as can
be seen from Table 2. The value of $\xi_2$ at $p$ turned out to be
$\xi_2(p)=45(5)$: the error is basically due to the uncertainty in the MC
values of $\kappa_2^c(T=0)$ and $\kappa_2^c(N_t=2)$.

We thus solved eq.(\ref{naive}) and eqs.(\ref{crit}) with respect to
$N_t^c$, using as inputs the values of $g_1^R$, $g_2^R$, $g^R$ and $\xi_1$
given in Table 2 and the above value of $\xi_2(p)$. In this way we obtained
the following estimates of $N_t^c$:
\be
N_t^{c({\rm one-loop})}=4.8(6)\;,\;\;\;\;\;\;\;N_t^{c({\rm gap})}=1.7(2).
\ee
It is clear that $N_t^{c({\rm one-loop})}$ is far from the MC value
$N_t=2$, while $N_t^{c(gap)}$ is quite close. It seems, then, that resummed
perturbation theory provides a rather accurate description of our data.

\section{Some comments about recent results on ISB on the Lattice}

Recently, there have been two other MC studies of ISB in four dimensions,
one by our group \cite{bim1} and another by Jansen and Laine \cite{jan}.
While in the first one no sign of ISB was found, the latter authors claimed
to have seen clear evidence of ISB and good agreement with resummed
perturbation theory. We would like to make here a few comments on these
works.

We shall start from our own previous work. The method followed there was
very similar to that used in this paper, but the simulations were performed
for a different choice of parameters. We think now that the essential
reason for the negative result is the fact that the conditions for ISB were
not enforced strongly enough. Near the simulation point, the absolute value
of $g^R$ was roughly the double of the value of $g_2^R$: according to the
one loop formulae (\ref{onel}), this was more than enough to induce ISB,
but nevertheless we did not find it. The reason of the failure is hidden in
the gap equations (\ref{gapeq}). Near our simulation point the correlation
length of $\phi_1$ was approximately one. This implies that, even at
highest temperature (corresponding to $N_t=2$) and so in the conditions
most favourable to ISB, $m_1/T=N_t/\xi_1$ was approximately equal to two.
By looking at the gap equations, it is easy to convince oneself that, with
a value of $m_1/T
\approx 2$, a $|g^R|$ which is twice $g_2^R$ is not enough to induce ISB.
The reason is the suppressing factor $h(x_1)$ in front of $g^R$ in the
second of eqs.(\ref{gapeq}): since $x_1\approx 2$, and $h(2)\approx 0.17$
it is clear that one would have needed a $|g^R|$ at least six times bigger
than $g_2^R$. In order to be on the safe side, in the new simulations
presented in this paper, since $\xi_1$ was still of order one, we selected
$|g^R|\approx 10 \;g_2^R$ which now was large enough.

As for ref.\cite{jan}, the authors searched for ISB in an $O(4) \times Z_2$
two scalar model. Their approach is very different from ours, and it is
useful to briefly review it here. Using perturbation theory (with two-loops
accuracy), they computed the relations giving the (bare) lattice
parameters, in terms of the renormalized masses and couplings, in the
symmetric phase of their model. Then, they performed a set of simulations
on lattices with $N_t=2,4$ (the few short runs on symmetric lattices,
according to the authors, should not be given much importance) for a number
of values of the bare couplings, corresponding to decreasing values for the
renormalized mass $m_1$ of the field expected to undergo ISB ($\phi_1$),
and constant values for the remaining renormalized parameters. In this way
they were able to vary the ratio $T/m_1$ in a large interval, even holding
$N_t$ fixed, or varying it very little, by just varying the renormalized
mass $m_1$. In the simulations they measured the expectation value of the
modulus of $\phi_1$, and the results turned out to be fully compatible with
the predictions of (resummed) perturbation theory. In particular, the field
$\phi_1$ appeared to be disordered for values of $T/m_1$ small enough, and
ordered for larger values of that ratio, which is a sign of ISB. Compared
with our approach, this procedure presents the advantage of avoiding the
very demanding task of an accurate determination of the critical point, but
on the other side it is necessarily limited to the region of very small
couplings, in order for perturbation theory to be applicable. Indeed, in
ref.\cite{jan}, perturbation theory was used not only to compute the
relations between the bare and renormalised couplings, as we said above,
but also to monitor the behavior of $<|\phi_1|>$ in the thermodynamical
limit. In view of the long lasting debate, whether ISB and SNR are genuine
effects or rather artifacts of perturbation theory, we preferred to pursue
the approach presented in this paper which, even if much more expensive, in
terms of computer time, does not make any recourse to perturbation theory.
As a reward, we were able to probe much larger values of the coupling
constants, something which is especially important in view of the
applications of ISB to realistic cosmological scenarios, where large
couplings in the scalar sector are usually needed in order to enforce ISB,
as was pointed out at the end of Sec.2. Moreover, we could also study in
detail the nature of the symmetry-breaking high-$T$ phase transition.

\section{Conclusions}

We have studied ISB at high temperature in four dimensions in a two-scalar
$\phi^4$ model with $Z_2 \times Z_2$ global symmetry. For fixed values of
the coupling constants $\lambda_1,
\lambda_2, \lambda$ and of the hopping parameter $\kappa_1$, we measured the critical
value of the hopping parameter $\kappa_2$, for which the field $\phi_2$
undergoes spontaneous symmetry breaking. This measure was performed both
at $T=0$ and at $T>0$. It was found that the value of $\kappa_2^c$ at $T>0$
is {\it smaller} than that for $T=0$, which is a signal of ISB. Due to the
extreme smallness of the shift, it was necessary to measure the values of
$\kappa_2^c$ with very high precision. This accuracy was reached via an
accurate analysis of FSS on large lattices with very high statistics. The
extreme difficulty of the measurements did not allow a scaling analysis of
the dependence of the result on the number of sites $N_t$ in the time
direction. For this reason, we cannot consider our results as conclusive.

In order to increase our confidence in the result, we compared the MC value
of the critical temperature $T_c$ with the theoretical value. While the
simple one-loop estimate turned out to be grossly incorrect, we found a
reasonable agreement with the value of $T_c$ obtained from the gap
equations (\ref{crit}), which result from the resummation of the ring
diagrams.

So, in conclusion, we have found some evidence that ISB is possible and
that resummed perturbation theory provides a qualitatively correct
description of the phenomenon. Nevertheless, we think that our results are
not conclusive and that further work is needed before ISB can be claimed
with confidence.


\end{document}